\documentclass[aps,twocolumn,showpacs,prl,reprint,nofootinbib]{revtex4-1}
\usepackage{epsfig}
\usepackage{dcolumn}
\usepackage{bm}
\usepackage{color}
\usepackage{ulem}
\usepackage{graphicx}
\usepackage{subfigure}
\usepackage{overpic}		
\usepackage{times} 
\usepackage[english]{babel}
\usepackage{amsmath}
\usepackage{amssymb}
\usepackage{mathrsfs}	
\usepackage{breqn}		
\usepackage{tabularx}
\addto{\captionsenglish}{%

}
\usepackage[colorlinks,linkcolor=blue,citecolor=blue]{hyperref}
\usepackage{lineno}

\newcommand{\psip}{\psi(3686)}

\newcommand{\XXb}{\Xi^{-}\bar{\Xi}^{+}}
\newcommand{\LLb}{\Lambda\bar{\Lambda}}
\newcommand{\pb}{\bar{p}}
\newcommand{\Xib}{\bar{\Xi}^{+}}
\newcommand{\Lamb}{\bar{\Lambda}}
\newcommand{\EE}{e^{+}e^{-}}

\newcommand{\Lm}{\mathcal{L}}
\newcommand{\W}{\mathcal{W}}
\newcommand{\Pm}{\mathcal{P}}
\newcommand{\N}{\mathcal{N}}
\newcommand{\III}{\uppercase\expandafter{\romannumeral3}}
\newcommand{\II}{\uppercase\expandafter{\romannumeral2}}
\newcommand{\alXib}{\alpha_{\bar{\Xi}^{+}}}
\newcommand{\alLamb}{\alpha_{\bar{\Lambda}}}
\newcommand{\phiXib}{\phi_{\bar{\Xi}^{+}}}
\newcommand{\AcpXi}{A_{CP, \Xi}}
\newcommand{\btable}{\begin{table}}
\newcommand{\etable}{\end{table}}
\newcommand{\btu}{\begin{tabular}}
\newcommand{\etu}{\end{tabular}}
\newcommand{\bfigure}{\begin{figure}}
\newcommand{\efigure}{\end{figure}}
\newcommand{\bitem}{\begin{itemize}}
\newcommand{\eitem}{\end{itemize}}
\newcommand{\bbt}{\bibitem}

\begin{document}
\normalsize
\parskip=5pt plus 1pt minus 1pt
\title{\boldmath
Observation of $\Xi^{-}$ hyperon transverse polarization in $\psip\to\XXb$}
\author{\small
M.~Ablikim$^{1}$, M.~N.~Achasov$^{10,b}$, P.~Adlarson$^{68}$, M.~Albrecht$^{4}$, R.~Aliberti$^{28}$, A.~Amoroso$^{67A,67C}$, M.~R.~An$^{32}$, Q.~An$^{64,50}$, X.~H.~Bai$^{58}$, Y.~Bai$^{49}$
% ...
\\
\vspace{0.2cm}
(BESIII Collaboration)\\
\vspace{0.2cm} {\it
$^{1}$ Institute of High Energy Physics, Beijing 100049, People's Republic of China\\
% ...
$^{4}$ Bochum Ruhr-University, D-44780 Bochum, Germany\\
% ...
$^{10}$ G.I. Budker Institute of Nuclear Physics SB RAS (BINP), Novosibirsk 630090, Russia\\
% ...
$^{28}$ Johannes Gutenberg University of Mainz, Johann-Joachim-Becher-Weg 45, D-55099 Mainz, Germany\\
% ...
$^{32}$ Liaoning Normal University, Dalian 116029, People's Republic of China\\
% ...
$^{49}$ Southeast University, Nanjing 211100, People's Republic of China\\
$^{50}$ State Key Laboratory of Particle Detection and Electronics, Beijing 100049, Hefei 230026, People's Republic of China\\
% ...
$^{58}$ University of Jinan, Jinan 250022, People's Republic of China\\
% ...
$^{64}$ University of Science and Technology of China, Hefei 230026, People's Republic of China\\
% ...
$^{67}$ University of Turin and INFN, (A)University of Turin, I-10125, Turin, Italy; (B)University of Eastern Piedmont, I-15121, Alessandria, Italy; (C)INFN, I-10125, Turin, Italy\\
$^{68}$ Uppsala University, Box 516, SE-75120 Uppsala, Sweden\\
% ...
$^{b}$ Also at the Novosibirsk State University, Novosibirsk, 630090, Russia\\
% ...
}
}
\date{\today}

\begin{abstract}
Using a sample of $(448.1~\pm~2.9)$ $\times 10^{6}$ $\psip$ decays collected with the BESIII detector at BEPCII, we report an observation of $\Xi^{-}$ transverse polarization with a significance of $7.3 \sigma$ in the decay $\psip\to\XXb$ {($\Xi^{-}\to{\Lambda\pi^-}$, $\Xib\to{\Lamb\pi^{+}}$, $\Lambda\to p\pi^{-}$, $\Lamb\to\pb\pi^{+}$)}. 
The relative phase of the electric and magnetic form factors is determined to be $\Delta\Phi = (0.667 \pm 0.111 \pm 0.058)$~rad. This is the first measurement of the relative phase for a $\psip$ decay into a pair of $\XXb$ hyperons. The $\Xi^{-}$ decay parameters ($\alpha_{\Xi^{-}}$, $\phi_{\Xi^-}$) and their conjugates ($\alXib$, $\phi_{\bar{\Xi}^{+}}$), the angular-distribution parameter $\alpha_{\psi}$, and the strong-phase difference $\delta_{p}-\delta_{s}$ for {$\Lambda\pi^-$} scattering are measured to be consistent with previous BESIII results.
\end{abstract}
\maketitle

Charmonium decays into hyperon and antihyperon pair offer a clean laboratory to explore hyperon  properties, such as their polarization and decay parameters, and to perform tests of fundamental symmetries. Taking advantage of the large $J/\psi$ and $\psi(3686)$ data samples and  the excellent performance of the detector,  BESIII has reported a series of observations of hyperon polarization in charmonium decays, such as  $J/\psi\to\LLb$~\cite{BESIII:2018cnd} and $J/\psi,\psip\to\Sigma^{+}\bar{\Sigma}^{-}$~\cite{BESIII:2020fqg}. This has led to a renewed interest in both theoretical and experimental investigations in hyperon physics. Baryon-spin polarization in a vector-charmonium decay  into a baryon-antibaryon pair was first considered in Ref.~\cite{Faldt:2017kgy}. Such an effect requires that the decay amplitudes are complex and have nontrivial relative phase $\Delta\Phi$.  To observe the baryon polarization some of the spin projections of the initial charmonium state must have different weights. This is possible if the state is produced in electron-positron annihilation with unpolarized beams since the photon, and therefore the charmonium, can only have $\pm 1$ helicities. A comparison of the polarization found in  $J/\psi$ and $\psip$ decays may elucidate other puzzles in vector charmonia decays, 
such as breaking of the `12\% rule', which predicts that the ratio of branching fractions of $\psi(3686)$ and $J/\psi$ decays into the same final state is 12\%~\cite{Appelquist:1974zd}. Experimentally, only one such polarization comparison has been performed, using the $J/\psi,\psip\to\Sigma^{+}\bar{\Sigma}^{-}$ reactions~\cite{BESIII:2020fqg}. The result reported in Ref.~\cite{BESIII:2020fqg} is striking since the relative phase between the form factors has opposite sign and different absolute values $\Delta\Phi_{J/\psi}=-0.270(12)$ rad and $\Delta\Phi_{\psip}={0.379(70)}$ rad. Currently, there is no explanation for this behavior. 
To improve understanding, measurements for other baryon-antibaryon pairs are needed.

\begin{figure}[!htbp]
			\centering
 			\includegraphics[width=0.4\textwidth]{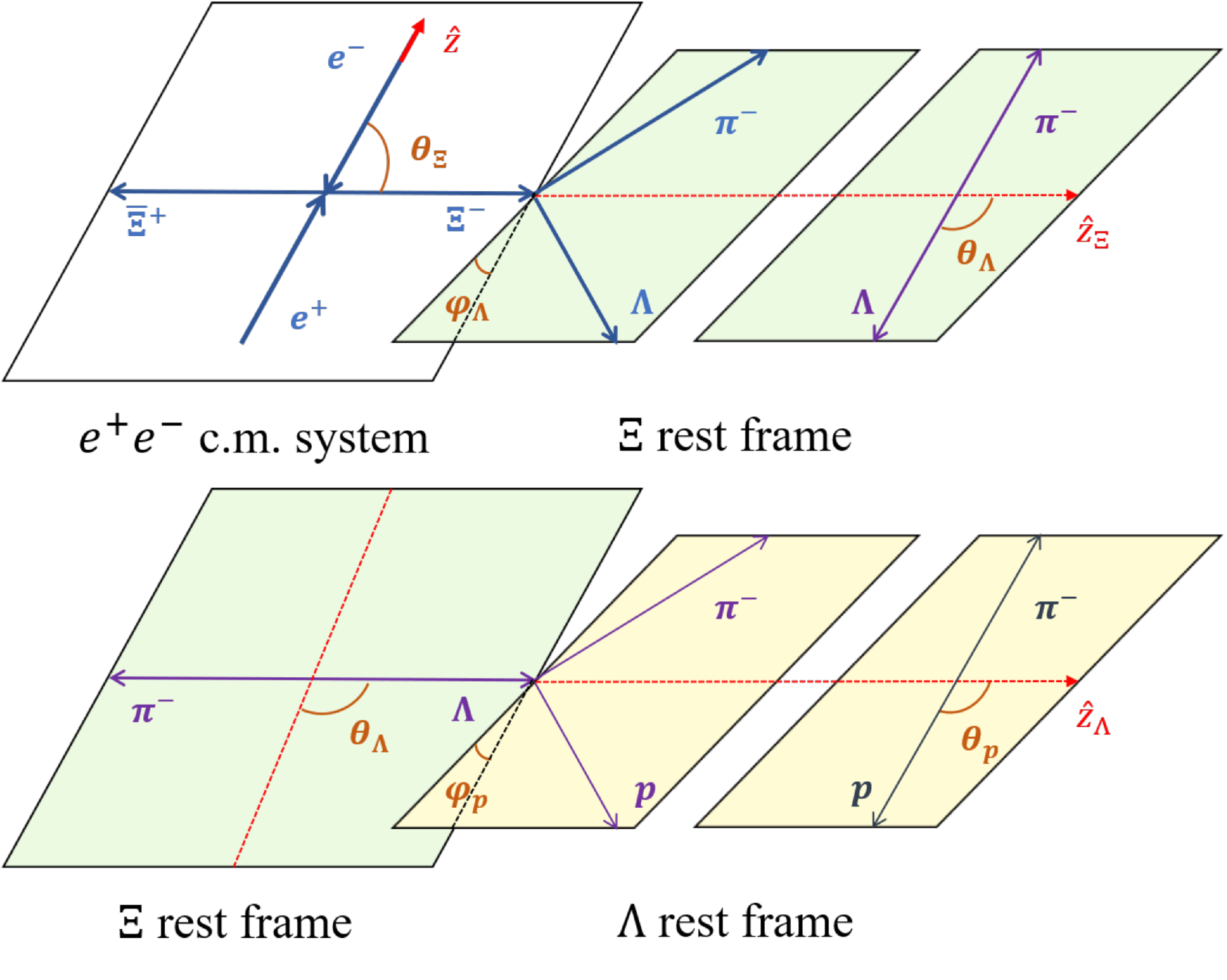}
			\caption{\small The definition of the helicity angles in different rest frames. The angles {$\theta_{\Xi},\theta_{\Lambda},\varphi_{\Lambda},\theta_{p},\varphi_{p}$} are the polar and azimuthal helicity angles of the $\Xi$, $\Lambda$ hyperons and the proton in three reference frames: $\EE$ {center-of-mass (c.m.) system}, $\Xi$ rest frame and $\Lambda$ rest frame, respectively. In the $\EE$ c.m. system, the $\hat{\boldsymbol z}$ axis is along the $e^+$ momentum direction, the $\hat{\boldsymbol z}_{\Xi}$ is along the $\Xi^-$ momentum direction. 
			In the $\Xi$ rest frame, the polar axis direction is $\hat{\boldsymbol z}_{\Xi}$, {$\hat{\boldsymbol y}_{\Xi}$ is along $\hat{\boldsymbol z}\times\hat{\boldsymbol z}_{\Xi}$} and $\hat{\boldsymbol z}_{\Lambda}$ is along the $\Lambda$ momentum direction. In the $\Lambda$ rest frame, the polar axis direction is $\hat{\boldsymbol z}_{\Lambda}$ and {$\hat{\boldsymbol y}_{\Lambda}$ is along $\hat{\boldsymbol z}_{\Xi}\times\hat{\boldsymbol z}_{\Lambda}$.} The vector ${\bf P}_{\Xi}\times\hat{\boldsymbol z}_{\Lambda}$ is along the $\hat{\boldsymbol y}_{\Lambda}$ axis.}
			\label{fig:orient}
		\end{figure}

A natural extension of these hyperon-polarization measurements is to investigate $\Xi^-$ decays in $J/\psi,\psip\to \XXb$, where the measurement of the polarization of the $\Xi^-$ hyperon is accessible via the process $\Xi^-\to{\Lambda\pi^-}$. The inclusion of charge-conjugate processes is implied throughout unless stated otherwise. The initial $\Xi$ transverse polarization with respect to the scattering plane, ${\bf P}_{\Xi}$, is related to the child $\Lambda$ polarizarion, ${\bf P}_{\Lambda}$, via the relation~\cite{Lee:1957qs},
    \begin{dmath}
        \label{eq:Pol}
            {\bf P}_\Lambda = \alpha_{\Xi^-} {\bf\hat z}_\Xi + \beta_{\Xi^-} {\bf P}_{\Xi} \times {\bf\hat z}_\Xi + \gamma_{\Xi^{-}} {\bf\hat z}_\Xi \times ({\bf P}_{\Xi} \times {\bf\hat z}_\Xi) ,\ 
    \end{dmath}
\noindent where $\hat{\boldsymbol z}_\Xi$ is the unit vector defined in Fig.~\ref{fig:orient} and $\alpha_{\Xi^{-}}$, $\beta_{\Xi^{-}}$, and $\gamma_{\Xi^{-}}$ are the asymmetry decay parameters for the $\Xi^-\to\Lambda\pi^-$ decay. The decay parameters $\beta_{\Xi^{-}}$ and $\gamma_{\Xi^{-}}$ are related and can be represented by a single parameter $\phi_{\Xi^-} = \tan^{-1}(\beta_{\Xi^-}/\gamma_{\Xi^-})$.
The joint angular distribution describes the spin correlation and polarization, which is expressed as{~\cite{Perotti:2018wxm}}
	\begin{dmath}
		\label{eq:W}
			\W(\boldsymbol\xi; \boldsymbol\Omega)
			= \sum^{3}_{\mu, \bar{\nu}=0}{\rm C}_{\mu\bar{\nu}}(\theta_{\Xi}; \alpha_{\psi}, \Delta{\Phi})\sum^{3}_{\mu'=0}\sum^{3}_{\nu'=0}a^{\Xi^-}_{\mu\mu'}a^{\Lambda}_{\mu'0}a^{\bar{\Xi}^+}_{\nu\nu'}a^{\Lamb}_{\nu'0}.
		\end{dmath}		
In Eq.~{\eqref{eq:W}}, ${\boldsymbol\Omega} = (\alpha_{\psi}, \Delta{\Phi}, \alpha_{\Lambda}, \alLamb, \alpha_{\Xi^{-}}, \alXib, \phi_{\Xi^{-}}, \phiXib)$ represent the production and decay parameters, the kinematic variables ${\boldsymbol\xi} = (\theta_{\Xi}, \theta_{\Lambda}, \phi_{\Lambda}, \theta_{\Lamb}, \varphi_{\Lamb}, \theta_{p}, \varphi_{p}, \theta_{\pb}, \varphi_{\pb})$ describe the production and the multistep decays, the spin density matrix ${\rm C}_{\mu\bar{\nu}}$ describes the spin configuration of the entangled hyperon-antihyperon pairs, and $a_{\mu\nu}^{H}$ is the propagation of the spin-density {matrix} based on the corresponding helicity variables and decay parameters.

Equation~\eqref{eq:W} characterizes the $\Xi^-$ polarization and spin correlation via the angular-distribution parameter $\alpha_\psi$ and $\Delta\Phi$. 
{The free parameters ($\boldsymbol\Omega$) are determined by a fit of the function defined in Eq.~\eqref{eq:W} to the data.}
If $\Delta\Phi$ is nonzero, it indicates a nonzero polarization of $\Xi^-$, and all other parameters can be determined simultaneously.

Recently BESIII observed $\Xi^-$ polarization in $J/\psi\to\XXb$ decays~\cite{BESIII:2021ypr} for the first time and the process was used for novel $CP$ symmetry tests by measuring the observables $A_{CP,\Xi} = (\alpha_{\Xi^-} + \alpha_{\bar \Xi^+})/ (\alpha_{\Xi^-} - \alpha_{\bar \Xi^+})$ and $\Delta\phi_{CP} = (\phi_{\Xi^-} + \phi_{\bar \Xi^+})/2$. 
{$CP$-violation tests performed with the $\psip\to\XXb$ data set have lower precision than those made in $J/\psi$ decays, but the sample can also be exploited for an independent measurement of {the strong phase difference, $\delta_p-\delta_s$, for {$\Lambda\pi^-$} scattering, where subscript $p(s)$ denotes P(S)-{wave}.} This quantity determines the sensitivity of the $A_{CP,\Xi}$ variable to the $CP$-violating phase difference $\xi_p-\xi_s$ in the decay. 
A nonzero value of $\delta_p-\delta_s$ is necessary to observe $CP$ violation via the $\AcpXi$ variable. It can be determined via the formula $\tan(\delta_p-\delta_s) = {\langle\beta_{\Xi}\rangle}/{\langle\alpha_{\Xi}\rangle}$. The value of $\langle\beta_{\Xi}\rangle$ is determined from the averages of decay parameters $\langle\alpha_{\Xi}\rangle$ and $\langle\phi_\Xi\rangle$. {We make these calculations of averages with method used in the particle data group (PDG) and take the same sign as decay parameters of $\Xi^-$ hyperon.} At present there is a tension between the $\phi_{\Xi^-}=-0.042\pm0.016$ rad {determined} from the HyperCP~\cite{HyperCP:2004not} experiment and the BESIII value for $\langle\phi_\Xi\rangle=0.016\pm0.014\pm0.007$ rad~\cite{BESIII:2021ypr}.} 
{This tension causes the difference of $\delta_p-\delta_s$ over one standard deviation in these two measurements, $\delta_p-\delta_s = (8.0\pm5.0)\times10^{-2}$ rad for HyperCP~\cite{HyperCP:2004not} and $\delta_p-\delta_s = (-4.0\pm3.3\pm1.7)\times10^{-2}$ rad for BESIII experiment~\cite{BESIII:2021ypr}.}
A model-dependent prediction of $CP$ violation within the context of the Standard Model (SM) gives $\AcpXi \sim -7\times10^{-5}$~\cite{Donoghue:1986hh}, and heavy-baryon perturbation theory predicts the strong-phase difference 
$\delta_p - \delta_s = (1.9 \pm 4.9)\times10^{-2}$ rad~\cite{Tandean:2002vy}.

In this paper we present an analysis of the angular distributions of the $\Xi^{-}$ hyperon and its final-state particles in the decay $\psip\to\XXb$, and subsequent $\Xi^-\to{\Lambda\pi^-}$ and $\Lambda\to p\pi^-$ decays, using a sample of $448.1 \times 10^6$ $\psip$ events~\cite{BESIII:2017tvm} collected with the BESIII detector~\cite{BESIII:2009fln}. In addition to measuring the spin polarization and decay asymmetry parameters of the $\Xi^{-}(\bar\Xi^{+})$ hyperon, the $CP$ observables $\AcpXi$ and $\Delta\phi_{CP}$ are also determined.

Candidates for $\psip\to\XXb$ events are selected using a full reconstruction of the subsequent decays $\Xi^-\to{\Lambda\pi^-}$, $\Lambda\to p\pi^-$. To study the decay parameters for the $\psip\to\XXb$ decay, five million Monte Carlo (MC) simulation events {with $\psip\to\XXb\to\pi^-\pi^+\LLb\to p\pb\pi^-\pi^-\pi^+\pi^+$} are generated according to a uniform distribution using the \textsc{kkmc} generator~\cite{Jadach:1999vf, Jadach:2000ir}. To study the potential background contributions, an inclusive MC sample of $350$ million $\psip$ decays is used, where the production of the $\psip$ resonance is simulated with the \textsc{kkmc} generator, the subsequent decays are processed via \textsc{evtgen}~\cite{Ping:2008zz,evt2-01} according to the measured branching fractions provided by the PDG~\cite{PDG}, and the remaining unmeasured decay modes are generated with \textsc{lundcharm}~\cite{lund, Yang:2014vra}.

Charged tracks are required to be reconstructed in the main drift chamber within its angular coverage $|\cos\theta|<0.93$, where $\theta$ is the polar angle with respect to the positron beam direction. 
Particle identification (PID) of the protons and pions is performed by considering the track momentum. Tracks with momentum above (below) 0.5~GeV/$c$ are assigned the proton (pion) hypothesis.
Events with at least two $\pi^-$, two $\pi^+$, one proton and one antiproton are retained for further analysis.

To reconstruct $\Lambda$ candidates, a secondary vertex fit~\cite{Xu:2009zzg} is applied to all $p\pi^{-}$ combinations. When there is more than one combination, {the one} with the minimum value of $\sqrt{(M_{p\pi^{-}} - {\rm m}_{\Lambda})^{2} + (M_{\bar{p}\pi^{+}} - {\rm m}_{\Lambda})^{2}}$ among all $p\pi^{-}$ combinations is chosen, where $M_{p\pi^{-}}$ is the invariant mass of the $p\pi^{-}$ pair and ${\rm m}_{\Lambda}$ is the known mass of the $\Lambda$ baryon~\cite{PDG}. The $p\pi^{-}$ invariant mass of the selected candidate is required to be within $5$ MeV$/c^{2}$ of the known $\Lambda$ mass. {This criterion is optimized by} the $\rm S/\sqrt{S + B}$ figure of merit (FOM) using MC simulation, where $\rm S$ is the number of signal MC events and $\rm B$ is the number of expected background events obtained from the simulation. To further suppress backgrounds from non-$\Lambda$ events, the $\Lambda$ decay length,
{\it i.e.} the distance between its production and decay positions, is required to be greater than zero.
The $\Xi^{-}$ candidates are reconstructed with a similar strategy using a secondary vertex fit and the candidate with the minimum value of $\sqrt{(M_{{\Lambda\pi^-}} - {\rm m}_{\Xi})^{2} + (M_{{\Lamb\pi^{+}}}-{\rm m}_\Xi)^{2}}$ from all ${\Lambda\pi^-}$ combinations is selected. Here, $M_{{\Lambda\pi^-}}$ is the invariant mass of the ${\Lambda\pi^-}$ pair and ${\rm m}_{\Xi}$ is the known mass of the $\Xi^{-}$ hyperon~\cite{PDG}. {Furthermore,} the $\Xi^{-}$ decay length
is required to be greater than zero.

\begin{figure}[!htbp]
 			\includegraphics[width=0.40\textwidth]{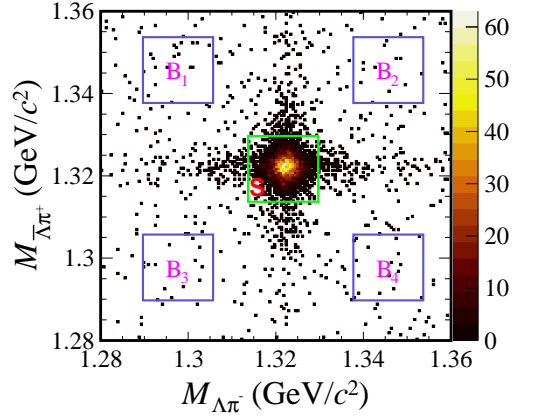}
			\caption{\small{Distribution of $M_{{\Lambda\pi^-}}$ versus $M_{{\Lamb\pi^{+}}}$. The box marked by S shows the signal region, the boxes marked by ${B}_j$ ($j=1, 2, 3, 4$) denote the sideband regions.}}
			\label{fig:sideband}
		\end{figure}

To further suppress backgrounds and improve the mass resolution, a four-constraint (4C) kinematic fit, imposing energy-momentum conservation from the initial $\EE$ to the final $\XXb$ state, is applied for all $\XXb$ {hypotheses} {after the full reconstruction}. Events with $\chi^{2}_{\rm 4C} < 200$ are retained based on the FOM {optimization}. 
Figure~\ref{fig:sideband} shows the two-dimensional distribution of $M_{{\Lambda\pi^-}}$ versus $M_{{\Lamb\pi^{+}}}$ {after performing the 4C kinematic fit}. A clear enhancement around the known $\Xi^{-}$ mass can be seen. The ${\Lambda\pi^-}$ invariant mass is required to be within $8$ MeV$/c^{2}$ of the known $\Xi^{-}$ mass, which is determined through the FOM. The signal region shown in Fig.~\ref{fig:sideband} is marked by S. After applying the above event selection criteria to the data, detailed event type analysis of the inclusive $\psip$ MC sample~\cite{Zhou:2020ksj} shows that the remaining backgrounds in this analysis are mainly from non-$\Xi^{-}$ events, such as $\EE\to\pi^{+}\pi^{-}\Lambda\bar\Lambda$. 
The background level is estimated using the sideband method. {We take four sideband regions and make a scaling of the backgrounds to the signal region, {\it i.e.}, $\sum^{4}_{j=1}{B}_{j}/4$}, where $j$ denotes one of the four regions shown in Fig.~\ref{fig:sideband}, and $B_j$ is the yield in that region. The upper and lower sideband regions are defined as $1.338$ GeV$/c^{2}<M_{{\Lambda\pi^-}}< 1.354$ GeV$/c^{2}$ and $1.290$ GeV$/c^{2}<M_{{\Lambda\pi^-}}< 1.306$ GeV$/c^{2}$ respectively, which means {that} the sideband region is about [6$\sigma$, 12$\sigma$] {away from the signal peak position}.
A total of $5358$ signal events survive the above selection criteria. {A scaled background of $15\pm4$ events} is estimated with the sideband method. A similar number of background events {($15\pm4$)} is estimated to be present from $\EE\to\XXb$ continuum production. This is deduced by applying the event selection to 2.9\,fb$^{-1}$ of data collected at a collision energy of 3.773~GeV/$c^2$~\cite{Ablikim:2013ntc, BESIII:2015equ}. In total therefore, the background level is determined to be $0.6$\%, which can be considered negligible for the analysis.

An unbinned maximum-likelihood {(MLL)} fit is performed to determine the $\Xi^{-}$ spin-polarization parameters. {Based on a previous study~\cite{BESIII:2018cnd}, we present $CP$ conservation in $\Lambda$ decays under the current statistics; thus, we assume $CP$ conservation of $\Lambda$ and fix the decay parameters $\alpha_{\Lambda/\bar\Lambda}$ to the average value $(\alpha_{\Lambda} - \alpha_{\bar\Lambda})/2$~\cite{BESIII:2018cnd} in the fit. This assumption would not affect the $CP$ tests in $\Xi$ decays under the current statistics.} The joint likelihood function is
\begin{equation}
	\label{eq:pdf2}
	\Lm
	=\prod^{N}_{i=1}\Pm(\boldsymbol\xi_{i};\boldsymbol\Omega)
	=\prod^{N}_{i=1}\frac{\W(\boldsymbol\xi_{i};\boldsymbol\Omega)\epsilon(\boldsymbol\xi_{i})}
	{\N(\boldsymbol\Omega)},
\end{equation}
{where $\Pm$ is the probability to produce event $i$ based on the measured parameters $\boldsymbol\xi_{i}$ and the set of observables $\boldsymbol\Omega$, $N$ is the number of data events after {all selection criteria}, 
$\W$ is defined in Eq.~\eqref{eq:W}, 
$\epsilon(\boldsymbol\xi_{i})$ is the detection efficiency and 
the normalization factor $ {\N(\boldsymbol\Omega)} = \int \W(\boldsymbol\xi; \boldsymbol\Omega) \epsilon(\boldsymbol\xi) d\boldsymbol\xi$.}
The negative value of the logarithm of $\Lm$ is minimized by using the \textsc{MINUIT} software package~\cite{James:1975dr}. The default fit contains no background term, as there is negligible contamination in the final sample. However, background is considered as a source of systematic uncertainty.

\begin{table*}[!htbp]
		\centering
		\caption{\small The numerical results for the measurements of the $\Xi^{-}$ polarization and the decay parameters. Also shown are the analogous parameters for the $J/\psi\to\XXb$ decay as reported in Ref.~\cite{BESIII:2021ypr}. 
		The first uncertainty is statistical and the second systematic.}
					\begin{tabular*}{\textwidth}{l@{\extracolsep{\fill}}rr}
				\hline \hline
				Parameter				 		        & \multicolumn{1}{c}{$\psip\to\XXb$}			& \multicolumn{1}{c}{$J/\psi\to\XXb$}\\	
				\hline
				$\alpha_{\psi}$ 				        & $0.693 \pm 0.048 \pm 0.049$	 	& $0.586 \pm 0.012 \pm 0.010$\\
				$\Delta{\Phi}$ (rad)				    & $0.667 \pm 0.111 \pm 0.058$	 	& $1.213 \pm 0.046 \pm 0.016$\\
				$\alpha_{\Xi^{-}}$			        	& $-0.344 \pm 0.025 \pm 0.007$	 	& $-0.376 \pm 0.007 \pm 0.003$\\
				$\alpha_{\bar{\Xi}^{+}}$			    & $0.355 \pm 0.025 \pm 0.002$	 	& $0.371 \pm 0.007 \pm 0.002$\\
				$\phi_{\Xi^{-}}$ (rad)				    & $0.023 \pm 0.074 \pm 0.003$	 	& $0.011 \pm 0.019 \pm 0.009$\\\vspace{0.5em}
 				$\phi_{\bar{\Xi}^{+}}$ (rad)			& $-0.123 \pm 0.073 \pm 0.004$	 	& $-0.021 \pm 0.019 \pm 0.007$\\ 
 				$\delta_{p}-\delta_{s}$ ($\times10^{-1}$ rad)		& $-2.0 \pm 1.3 \pm 0.1$		& $-0.40 \pm 0.33 \pm 0.17$\\
				$\AcpXi$ ($\times10^{-2}$) 						    & $-1.5 \pm 5.1 \pm 1.0$		& $0.60 \pm 1.34 \pm 0.56$ \\
				$\Delta\phi_{CP}$ ($\times10^{-2}$rad)		        & $-5.0 \pm 5.2 \pm 0.3$		& $-0.48 \pm 1.37 \pm 0.29$ \\
				\hline \hline
			\end{tabular*}
		\label{tab:paras}
		\end{table*}

The results of the fit are given in Table~\ref{tab:paras}. The relative phase $\Delta{\Phi}$ for $\psip\to\XXb$ decay is measured for the first time and differs from zero. The decay-asymmetry parameters $\alpha_{\Xi^{-}}$ and $\alpha_{\Xi^{+}}$ agree within their statistical uncertainties but are somewhat lower than the PDG average value ($\alpha_{\Xi^{-}}$ = $-$0.401 $\pm$ 0.010) \cite{PDG}.  
The strong-phase difference, $\delta_{p}-\delta_{s}$, is consistent with the BESIII result measured from the $J/\psi \to \XXb$ decay~\cite{BESIII:2021ypr}, and lies around two standard deviations below the HyperCP measurement~\cite{HyperCP:2004not}.
The other parameters are compatible with earlier measurements.

The transverse $\Xi^{-}$ polarization, $P_y$, is given by
\begin{dmath}
    \label{eq:Py}
        \mathrm{P}_y = \sqrt{1-\alpha^2_\psi}\sin\theta_{\Xi}\cos\theta_{\Xi}\sin\Delta\Phi/(1+\alpha_\psi\cos^2\!\theta_{\Xi})
\end{dmath}
and the spin correlations ${\rm C}_{ij}$ ($i, j = x, y, z$),
are related to the {${\rm C}_{\mu\bar\nu}$} in Eq.~\eqref{eq:W} as follows

\begin{figure}[!htbp]
			\centering
			\subfigure{\includegraphics[width = 0.23\textwidth]{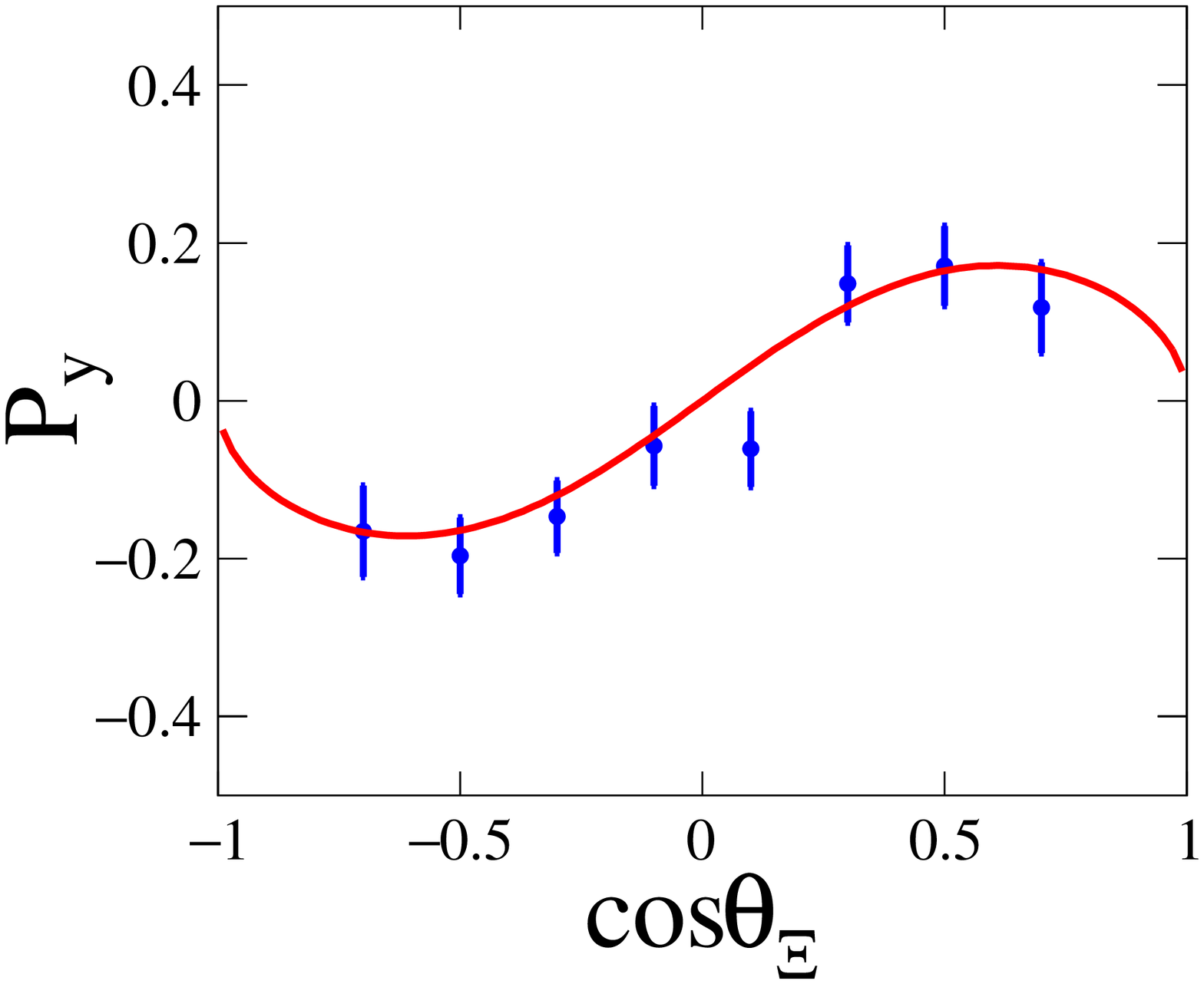}}
			\subfigure{\includegraphics[width = 0.23\textwidth]{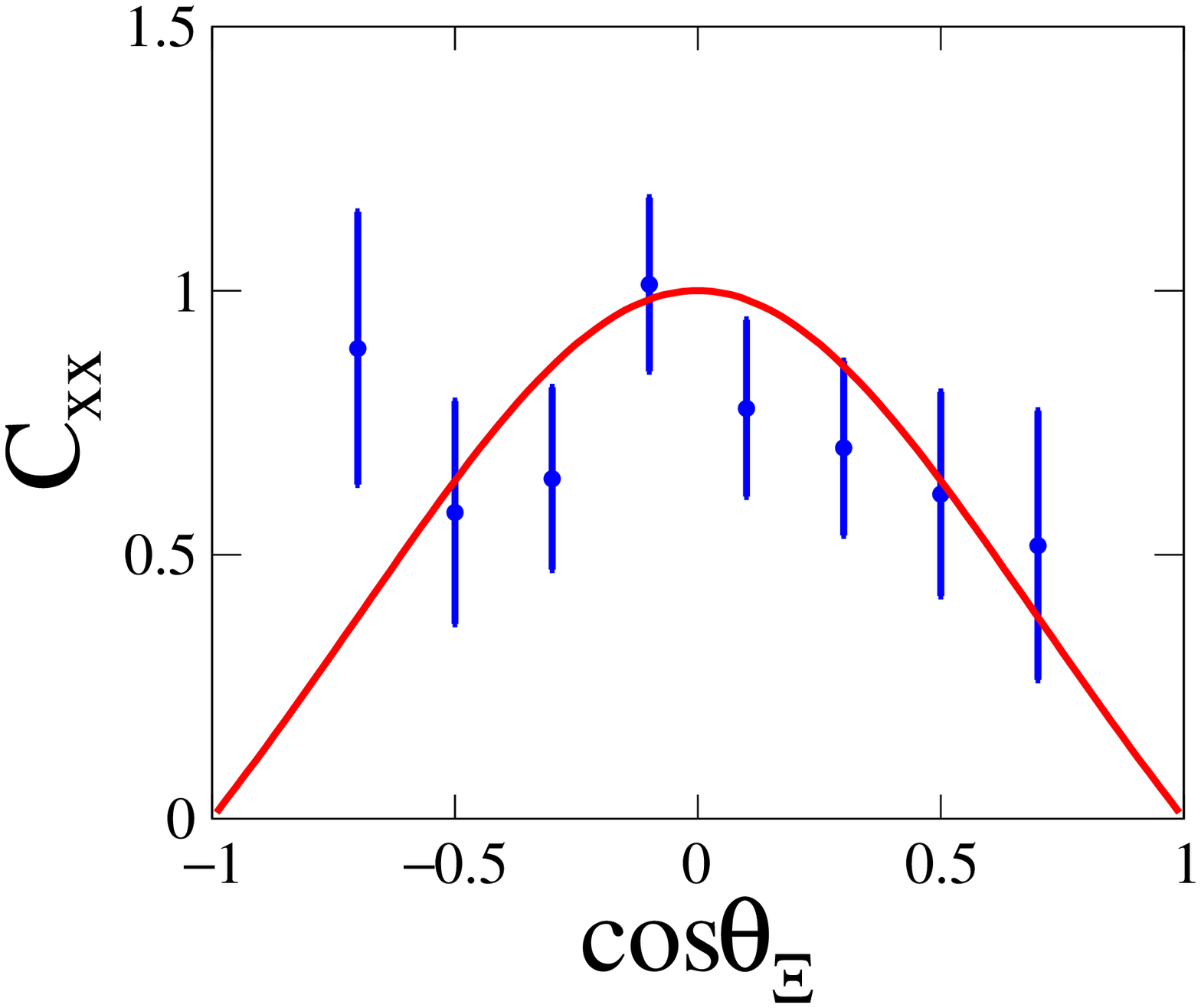}} \\
			\subfigure{\includegraphics[width = 0.23\textwidth]{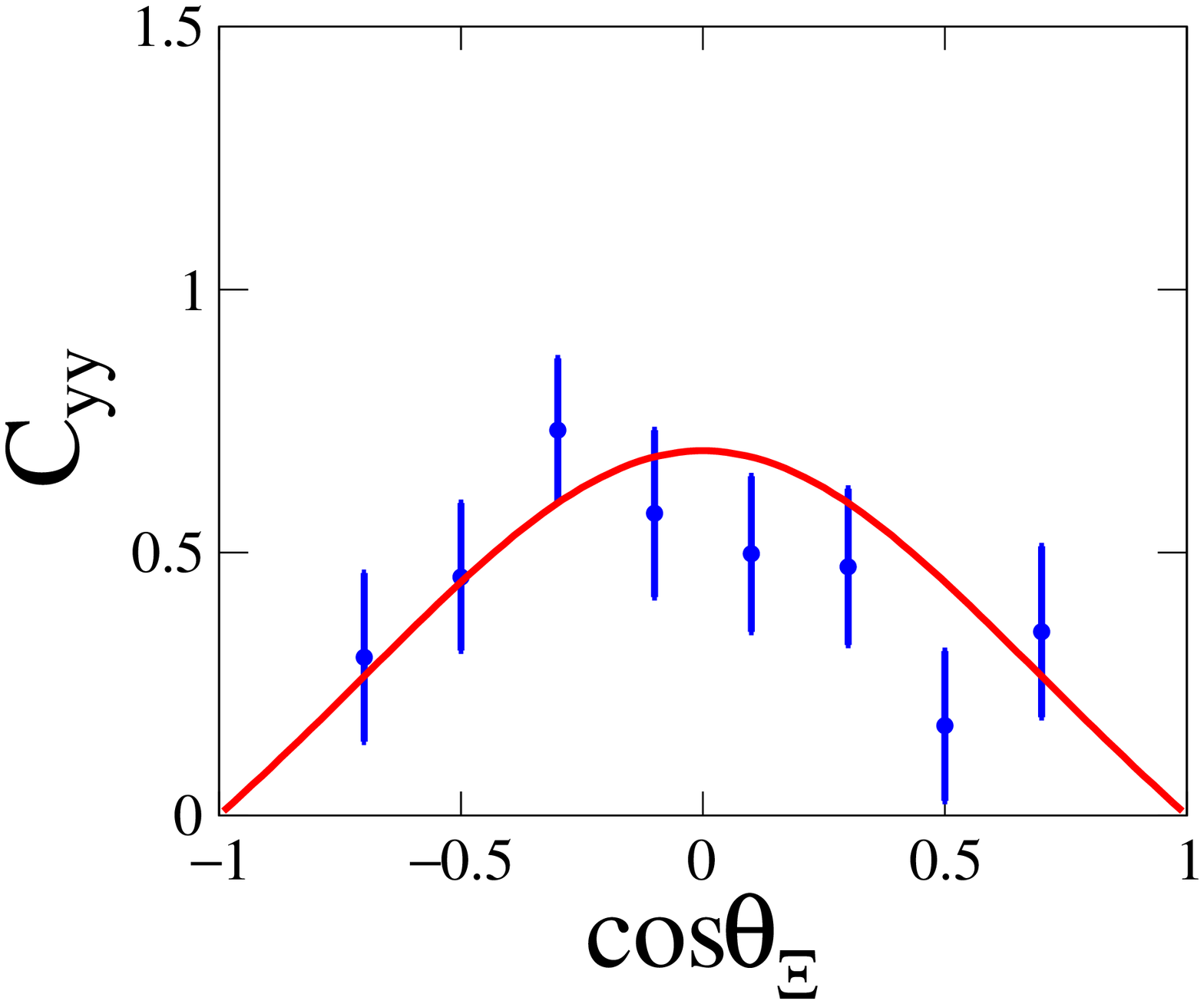}}
			\subfigure{\includegraphics[width = 0.23\textwidth]{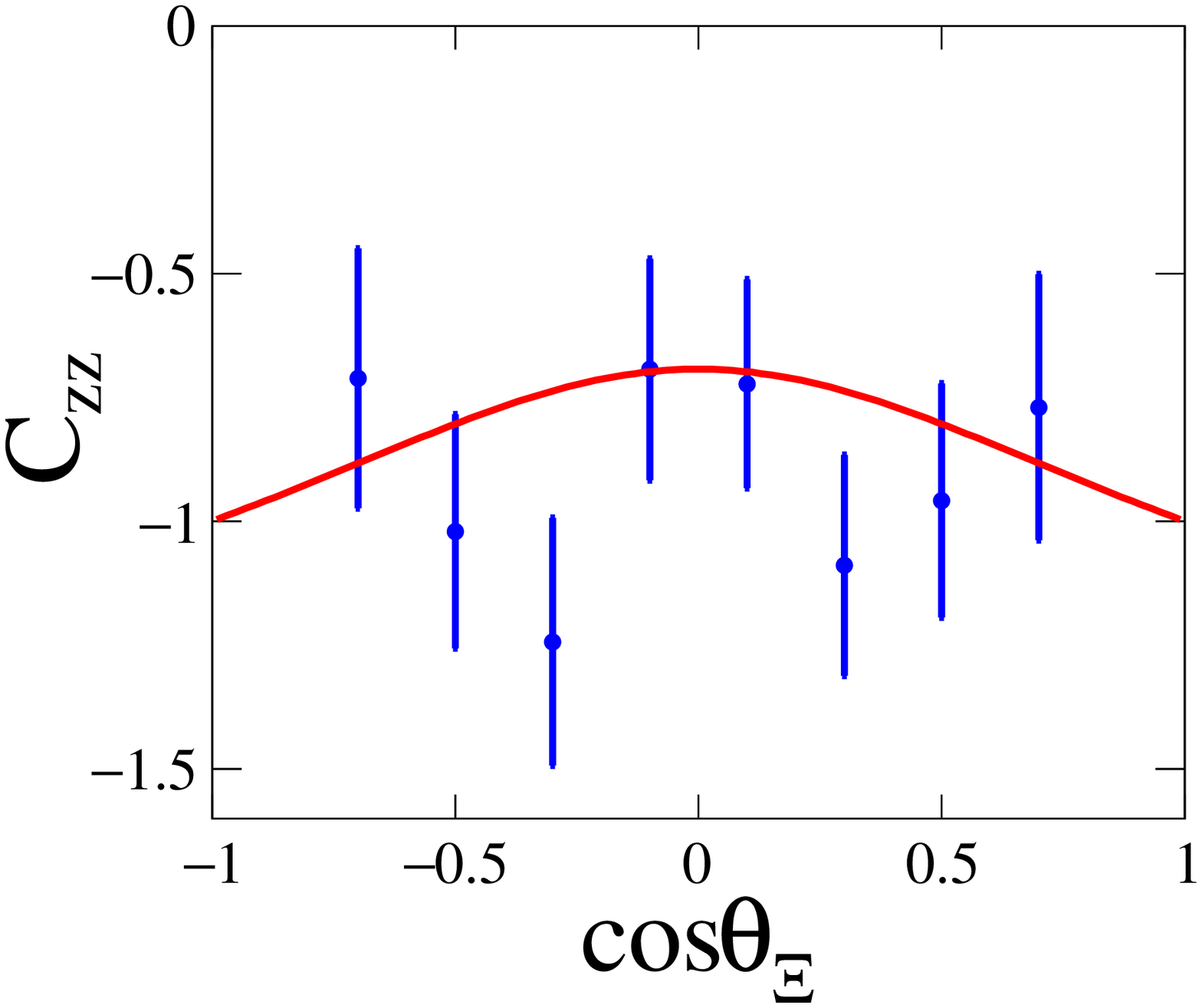}}
			\caption{\small Polarization $P_y$ and spin correlations ${\rm C}_{xx}$, ${\rm C}_{yy}$ and ${\rm C}_{zz}$ in $\psip\to\XXb$ as a function of $\cos\theta_{\Xi}$. The dots with error bars are the binned experimental result and the red lines give the result of the {global} MLL fit. {The error bar represents the statistical uncertainty.}}
			\label{fig:moment}
		\end{figure}

\begin{dmath}
		\label{eq:mu}
			{{\rm C}_{\mu\bar\nu}} = (1+\alpha_{\psi}\cos^2\!\theta_{\Xi})
			\begin{pmatrix}
			1	&0		&P_y	&0\\
			0	&{\rm C}_{xx}	&0		&{\rm C}_{xz}\\
			-P_y	&0		&{\rm C}_{yy}	&0\\
			0	&-{\rm C}_{xz}	&0		&{\rm C}_{zz}
			\end{pmatrix}.
		\end{dmath}
Figure~\ref{fig:moment} shows the resulting $P_y$ polarization and spin correlations together with the fit results. The $P_y$ polarization and the spin correlations are clearly nonzero.

Possible sources of systematic bias on the parameter measurements are considered. They include the fit method, the $\Xi^{-}$ reconstruction, the 4C kinematic fit, the background and the decay parameters for $\Lambda\to p\pi$.
We validate the reliability of the fit results by an input and output check based on $100$ pseudodata samples. The helicity amplitude formula is based on Eq.~\eqref{eq:W} and the measured polarization parameters in this work are used as the input parameters, where the central values used in the generation of the number of events for each pseudodata sample are the same as measured in the data sample. The average difference between input and the output result is taken as the systematic uncertainty.
The systematic uncertainty due to the $\Xi^{-}$ hyperon reconstruction incorporates the track selection, PID criteria, requirements on mass window and decay length of $\Xi^{-}(\Lambda)$, and is studied with a control sample of $\psip\to\XXb$ decays using the same method as in Refs.~\cite{BESIII:2012ghz, BESIII:2016ssr, BESIII:2016nix, BESIII:2019dve, BESIII:2019cuv, BESIII:2021ccp, BESIII:2020ktn, BESIII:2021aer, BESIII:2021cvv}. From this we obtain the efficiency difference between the data and MC event reconstruction. Then we apply a correction to the MC simulation with the efficiency difference described above, and repeat the fit. The differences between the new values and the nominal values are taken as the systematic uncertainties of the parameters resulting from the $\Xi^{-}$ reconstruction.
The systematic uncertainty due to the 4C kinematic fit is studied by correcting the helical-track parameters for each charged particle. For the nominal result, we take the uncorrected results. We repeat the full fit procedure using a MC sample with the track correction and take the difference between the two fit results as a systematic uncertainty. 
To assign a systematic uncertainty from background, which is not included in the nominal fit, we repeat the fit taking in consideration the backgrounds from the sideband region and continuum process in the log-likelihood function.  The differences between the new values and the nominal values are taken as the systematic uncertainties of the parameters resulting from backgrounds.
The uncertainty from the decay parameters $\alpha_{\Lambda/\bar\Lambda}$ of $\Lambda\to p\pi^{-}/\Lamb\to\pb\pi^+$ is estimated by varying the nominal value, obtained by the average in Ref.~\cite{BESIII:2018cnd}, by $\pm 1 \sigma$. The largest difference in the results is taken as the systematic uncertainty.
Assuming all sources are independent,  {the total systematic uncertainty is} determined by taking the square root of their the quadratic sum and the {results} are given in Table~\ref{tab.sumSU}.

	\begin{table*}[!htbp]
		\caption{\small The {absolute} systematic uncertainties on the fit parameters. {(Entries of 0.000 indicate a contribution less than 0.001.)}}
		\centering
			\begin{tabular*}{\textwidth}{l@{\extracolsep{\fill}}cccccc}
				\hline \hline
				Source                                	& $\alpha_{\psi}$ & $\Delta{\Phi}$ & $\alpha_{\Xi^{-}}$ & $\alpha_{\bar{\Xi}^{+}}$  &  $\phi_{\Xi^{-}}$ & $\phi_{\bar{\Xi}^{+}}$ \\
				\hline
				Fit method                    			& $0.001$         	& $0.026$         	& $0.006$    	& $0.000$     	& $0.002$                    & $0.001$                  \\
				$\Xi^{-}$ reconstruction		& $0.027$		& $0.027$		& $0.000$	& $0.000$		& $0.001$			& $0.000$ \\
				$\bar\Xi^{+}$ reconstruction	& $0.040$		& $0.042$		& $0.001$	& $0.000$		& $0.001$			& $0.000$ \\
				4C kinematic fit              		& $0.002$         	& $0.004$          	& $0.002$    	& $0.001$    	& $0.002$                    & $0.004$                  \\
				Sideband 			  	              	& $0.004$         	& $0.004$          	& $0.000$     	& $0.000$     	& $0.000$                    & $0.000$                  \\
				Continuum process 			      	& $0.002$         	& $0.001$          	& $0.000$     	& $0.000$      	& $0.000$                    & $0.000$                  \\\vspace{0.5em}
				$\Lambda$-decay parameter   	& $0.006$         	& $0.011$          	& $0.001$    	& $0.002$      	& $0.000$                    & $0.000$                  \\
				Total                         		      		& $0.049$         	& $0.058$          	& $0.007$     	& $0.002$     	& $0.003$                    & $0.004$     		\\
				\hline \hline
			\end{tabular*}
		\label{tab.sumSU}	
		\end{table*}

In summary, we report an observation of the $\Xi^-$ hyperon spin polarization in the process $e^{+}e^{-}\to\psip\to\XXb$.
The measured parameters  are summarized in Table~\ref{tab:paras}. The relative phase between the electric and magnetic psionic form factors, $\Delta\Phi$, in the $\psip\to\XXb$ decay is determined to be different from zero for the first time with a significance of $7.3 \sigma$, including the systematic uncertainty, by comparing the most conservative differences of likelihoods of the baseline fit and the one assuming no polarization.
{This result differs significantly from the corresponding relative phase in the $J/\psi$ decay~\cite{BESIII:2021ypr}. The behavior is also noticeably different to that seen in decays of charmonia to $\Sigma$ baryons, where the corresponding phase is negative and around four times smaller in magnitude than for $\Xi$ baryons in $J/\psi$ decays, and around two times smaller in the case of $\psip$ decays~\cite{BESIII:2020fqg}.
The results provide important input for understanding the difference of the $\alpha_{\psi}$, $\Delta\Phi$ values at different energies and different charmonia states, and even insight into double-strange versus single-strange hyperon pair production~\cite{BaldiniFerroli:2019abd, Ferroli:2020mra}.
Further input from the theory community and new measurements in related decay modes are needed to interpret these results. 
Other decay parameters ($\alpha_{\Xi^{-}}$, $\alXib$, $\phi_{\Xi}$, $\phi_{\bar{\Xi}^{+}}$}) for $\Xi^-\to{\Lambda\pi^-}$ and $\Xib\to{\Lamb\pi^{+}}$ decays including the angular distribution parameters ($\alpha_{\psi}$) are measured to be consistent with previous publications within their uncertainties~\cite{BESIII:2021ypr,BESIII:2016ssr}.
The measurement of the strong phase difference, $\delta_p-\delta_s$, constitutes an independent determination of this important quantity describing the strong interaction between the $\Lambda$ and $\pi^-$ hadrons. The result is consistent with the heavy-baryon perturbation theory \cite{Tandean:2002vy} and agrees with the recent BESIII measurement performed with $J/\psi$ decays~\cite{BESIII:2021ypr}, and is around two standard deviations lower than the HyperCP measurement of this parameter~\cite{HyperCP:2004not}.
The $CP$ observables $\AcpXi$ and $\Delta\phi_{CP}$ are determined from the measurements of the $\Xi^-(\Xib)$ decay parameters. The results are consistent with the $CP$ conservation hypothesis. Our measurement of $\AcpXi$ is in agreement with the SM expectation.

\section{Acknowledgments}
The BESIII Collaboration thanks the staff of BEPCII and the IHEP computing center for their strong support. This work is supported in part by National Key R\&D Program of China under Contracts Nos. 2020YFA0406400, 2020YFA0406300; National Natural Science Foundation of China (NSFC) under Contracts Nos. 12075107, 11625523, 11635010, 11735014, 11822506, 11835012, 11905236, 11935015, 11935016, 11935018, 11961141012, 12047501, 12022510, 12025502, 12035009, 12035013, 12061131003 and 12225509; 
the Chinese Academy of Sciences (CAS) Large-Scale Scientific Facility Program; Joint Large-Scale Scientific Facility Funds of the NSFC and CAS under Contracts Nos. U1732263, U1832207; CAS Key Research Program of Frontier Sciences under Contract No. QYZDJ-SSW-SLH040; 100 Talents Program of CAS; Institute of Nuclear and Particle Physics, Astronomy and Cosmology (INPAC) and Shanghai Key Laboratory for Particle Physics and Cosmology; ERC under Contract No. 758462; European Union Horizon 2020 research and innovation programme under Contract No. Marie Sklodowska-Curie grant agreement No 894790; German Research Foundation DFG under Contracts Nos. 443159800, Collaborative Research Center CRC 1044, FOR 2359, GRK 2149; Istituto Nazionale di Fisica Nucleare, Italy; Ministry of Development of Turkey under Contract No. DPT2006K-120470; National Science and Technology fund; Olle Engkvist Foundation under Contract No. 200-0605; Lundstr{\"o}m-${\rm \mathring A}$mans Foundation; STFC (United Kingdom); The Knut and Alice Wallenberg Foundation (Sweden) under Contract No. 2016.0157; The Royal Society, (United Kingdom) under Contracts Nos. DH140054, DH160214; The Swedish Research Council; Polish National Science Centre under Contract No. 2019/35/O/ST2/02907; U. S. Department of Energy under Contracts Nos. DE-FG02-05ER41374, DE-SC-0012069.

\end{document}